# Solid-to-super Critical Phase Change during Laser Internal Ablation

Yan Li[a,b,*], Chong Li[b], Chao Yang[a]

[a]Department of Mechanical Engineering, Ocean University of China,

Qingdao 266100, P.R. China

[b]Department of Mechanical Engineering, Iowa State University,

Ames, IA 50011, USA

## Abstract

The mechanisms of phase change of argon during picosecond laser internal ablation are studied using molecular dynamics simulations. It is found that propagation of stress wave and fluctuation of temperature are periodical. The phase change process from solid to liquid to supercritical fluid then back to solid occurs as combined results of heating and the propagation of tensile stress wave induced by the laser pulse and the limited internal space.

**Keywords:** Laser-induced, Molecular dynamics, Temperature fluctuation, Super critical fluid, Phase change

---

[*] Corresponding author. Tel: +86 13953259329, Fax: +86 532 66781550, Email: yanli@ouc.edu.cn

# 1. Introduction

Pulsed laser ablation is the process of material removal and processing after the target is irradiated by intensive laser pulses. It is well documented that pulses with very short durations, such as picosecond or femtosecond, are advantageous in many applications. Picosecond laser ablation has become one of the most intensively investigated topics in the research of laser material interaction. The basic mechanism has been studied in previous Molecular dynamic (MD) works. But the internal laser ablation study in MD is quite new. Picosecond laser ablation occurs at very short temporal and spatial scales, involving complicated optical, thermodynamic, energy transfer, and mechanical processes which are closely coupled. At the same time, the target could be heated to extremely high temperature and pressure, where thermal and mechanical properties of the material are generally unknown [1]. For high fluence and short pulse, the target is unable to boil, and the near-surface region relaxes explosively into a mixture of vapor and equilibrium liquid droplets, which is called phase explosion [2]. Song and Xu [3] used nickel as the specimen to study the laser fluence threshold of phase explosion. It is possible to directly heat the solid material to be above the critical temperature, followed by expansion leading to the thermodynamically unstable region, causing material decomposition [4]. This material decomposition process, from solid to supercritical fluid, then to the un-stable region is termed critical point phase separation [1].

Many groups have conducted experimental and numerical research on the phase change process during nanosecond laser ablation [3, 5-10]. Information such as the transient temperature during the laser ablation process, the creation of the vapor phase inside the superheated liquid, and the time required to form a two-phase mixture were studied for the phase change phenomena induced by a femtosecond laser pulse [11].

Nanostructing is a typical application of laser-material interaction. Nanostructure can be formed inside photosensitive glass using a focused femtosecond laser [12, 13]. The discovery of long-range, self-organized, periodic planar nanocrack structures in fused silica is an important development in the field of laser dielectric modification [14]. Techniques for two dimensional (2D) or three dimensional (3D) image engraving inside crystal have been developed utilizing its transparency and high refractive index [15, 16]. Ionization due to multiphoton and avalanche increases absorption in the localized area of optical breakdown [17]. Thanks to that absorption of laser energy, matter inside the breakdown area is heated and the chain of processes considered in the analyzed paper is triggered.

The internal ablation process is very difficult to study since a lot of experimental techniques cannot reach the material inside to measure the temperature, stress and so on. There are experiments and computer simulations done by French group [18] and Bulgakova [19] where 3D pattern of developing of optical breakdown in transparent solid has been considered. In experiments they follow on the subpicosecond evolution of breakdown development by using pump-probe technique. In addition, internal ablation has very different characteristics from the open-space ablation: the material's phase change is extremely confined by the solid material surrounding it. The constraint in the open space has been studied in Wang's group [20-22]. Laser-material interaction with the presence of ambient gas leads to the formation of shock wave, which can significantly affect the phase change process. Modeling work about plume propagation in vacuum and background gas has been reported [23]. Although the interaction of the plume with ambient gas significantly suppresses the void formation and phase explosion, no obvious effect is found on the stress wave within the target. Very interestingly, secondary stress waves resulting from re-deposition of ablated atoms and void collapse are observed, although their magnitude is much smaller than the directly laser-induced stress wave [20]. The spatial confinement-induced temperature rise has been reported [21]. Furthermore, the effect of SPM tip confinement in the nanostructuring has been studied [22]. The dynamics of the melting of a surface nanolayer and the formation of thermal and shock waves in metals irradiated by femtosecond laser pulses has been investigated both experimentally and theoretically[24]. Micro-explosion in a confined domain results in a sub-micron cavity formation. The numerical simulations show the cavity size is strongly dependent on the parameters of the equation of state such as the Grüneisen coefficient or the latent heat of sublimation[25]. However, the solid-to-super critical phase change during the internal ablation confined by the target has not been discussed yet.

In this work, molecular dynamics (MD) simulation is employed to investigate picosecond laser ablation inside an argon crystal of 3,200,000 atoms, with an emphasis on the understanding of the mechanism of phase change from solid to supercritical fluid during laser ablation inside the target. Laser induced heat transfer, stress wave, phase change, and material ablation are studied. The material undergoes an extremely confined phase change process. Its maximum temperature can be above the thermodynamic critical point. The purpose of this work is to use numerical techniques to investigate the rapid phase change process and thermodynamic trajectories of atoms during picosecond laser internal ablation. Parameters relevant to laser ablation, such as the temperature, stress and volume in the laser ablation process are studied in detail.

## 2. Physical Domain Construction and Modeling

Figure 1(a) shows the physical model of the simulation and the domain construction. The target measures 108.28 nm×10.82 nm×108.28 nm (x×y×z). The thickness of the whole simulation system in the *y* direction is 10.82 nm: relatively small compared with the *x* and *z* direction size. The quasi-2D design makes it is possible to study the phase change and stress waves to a very long time. Periodic boundary conditions are applied in the *x* and *y* directions. The boundary conditions in the *z* direction are the absorptive boundary condition at the bottom and shrink boundary condition on the top. An effective absorptive boundary condition should accurately simulate the wave radiation into the exterior, while keeping the computational cost low [26-28]. When a laser pulse irradiates the material from the top side, a strong stress wave will emerge in the material and propagate to the bottom side. When this strong stress wave reaches the bottom boundary of the MD domain, the strong tensile stress can tear the material near the bottom boundary of the MD domain, inducing unrealistic material damage. The absorptive boundary treatment has been proved to work very well in terms of eliminating stress wave reflection and avoiding undesired material damage in the boundary region. A region of 10 Å along the *z* axis at the bottom of the target is chosen as the absorptive boundary. An external force specified by the work [27, 29] is added to the atoms in this region. For shrink boundary condition, the position of the face is set so as to encompass the atoms in that dimension (shrink-wrapping), no matter how far they move. The upper boundary is shrink boundary to see the shift when the atoms move out of the surface. The interaction between atoms follows the Lennard-Jones (12-6) potential. The laser irradiation is focused on a circular spot with a radius of 2 nm in the target. Figure 1(b) shows the laser beam intensity distribution. Of course, the real illuminated spot sizes are at least of the order of optical wavelength that is of the order of micron. The radius 2 nm of the disk used as absorptive body in the presented simulations can reduce the atom number but show the mechanism of the internal ablation. A single laser pulse of 40 ps is applied in this simulation work. The laser energy is spatially uniform over the spot. The full width at half maximum (FWHM) of the incident laser beam intensity distribution is 11.5 ps and the peak occurs at $t=$ 9 ps. Laser energy absorption obeys the Beer-Lambert law. To accomplish it, the laser absorption region is divided into a number of cells of thickness $\Delta z=$ 1 nm. An artificial absorption depth of 5 nm is used in our work. The incident laser energy decreases exponentially after the absorption in each cell. The details have been documented in the previous works [20, 29, 30].

After constructing the whole physical domain for modeling and definition of atom interaction and boundary conditions, the system is first treated as a canonical ensemble (NVT) and modeled for 1.25 ns and then a microcanonical ensemble (NVE) for 500 ps to reach thermal equilibrium. Then the laser energy $E=$ 10 J/m$^2$ is applied inside the target. After that, the whole system is in NVE again for recrystallization. During the laser irradiation and in the whole stage of laser ablation, the time step is set to 2 fs. For MD calculations, due to the limitation of computing power, the number of atoms in the system has been restricted. Argon is chosen as the material in this simulation work, due to its great computational efficiency. This is critical for this work since our modeled system is composed of a large number of atoms (3,200,000 atoms) and the whole physical process under simulation lasts for a long time. The argon crystal is arranged in the simple face-centered cubic structure and well described by the Lennard-Jones (12-6) potential. It is easy for the modeling and fast for the computation while the conclusion does not lose its generality. Its validity has been proved in laser-material interaction simulation [20, 29-33]. The laser absorption process depends on the laser wavelength and the light absorption characteristics of material. Absorption depth $\tau$ is usually calculated as: $\tau = \lambda / (4\pi \cdot K)$. $\lambda$ is the laser absorption depth. $K$ is the material imaginary reflection index and it is the unique laser light absorption characteristic of material argon. The LJ potential well depth $\varepsilon$ is 0.0103 ev and the equilibrium separation parameter $\sigma$ is 3.406 Å. And the cut-off distance is set to $2.5\sigma$. LAMMPS is employed in this work [34].

## 3. Structure and Phase Change: A General Picture

Internal nanostructuring is affected by the incident laser and the target material properties. A case of laser fluence $E= 10$ J/m$^2$ is chosen to demonstrate the general pictures of laser ablation and recrystallization process. Figure 2 is the atomic snapshots at different times. Laser energy is applied from $t= 0$ ps to $t= 40$ ps. It is clearly shown that the ablation has already started at $t= 5$ ps. The dark region in the target indicates the hot and amorphous status of the small part in the target. And some destruction is permanent even the region under destruction shrinks at last compared to the initial moment after the laser pulse. No voids are observed within the material.

Even though the laser pulse is very short, the thermal expansion lasts a relatively long time after the laser pulse. Figures from $t= 5$ ps to $t= 100$ ps show the thermal expansion and relaxation of the target. From $t= 200$ ps to the end of the simulation $t= 1460$ ps, recrystallization is observed. Because of the strong stress wave in the target, the structure of the material is distorted from the well-arranged crystal structure as shown at $t= 5$ ps. From the atomic configuration, it is seen that a lot of the damages are temporary and they are back to the good crystal structure again in the recrystallization process. However, the destruction in the laser irradiation region is permanent. At the end of simulation at $t= 1460$ ps, defects in the target are still observed.

Figure 3 is for the stress contour of the system. The stress wave in the target is very clear. Our previous work has presented detailed analysis and discussion of the stress wave in a solid [31, 35] under laser ablation. The stress wave forms at 10 ps and propagates in the *x* and *z* directions. The stress refers to the normal stress and is calculated as $\sigma = (\sigma_{xx} + \sigma_{yy} + \sigma_{zz})/3$. The stress moves from the center to the boundary at the beginning. Before *t*= 50 ps, the stress wave spreading from the center is observed. The stress wave re-enter from the opposite side to the center at around *t*= 90-100 ps, which is obvious in the *x* direction because of the boundary condition. The re-entered stress wave shows up at around *t*=180 ps and *t*=260 ps again. This indicates the process of stress propagation is periodic with a period about 80 ps that also can be seen in Fig.6. The velocity of the stress wave is smaller than but is close to the local sound speed (1275 m/s) in Argon [36].

Figure 4 is for the temperature contours at different times. Melting temperature of argon crystal at normal conditions is 83.95K. Initially, the thermal equilibrium at *T*= 50 K of the whole system is achieved before the laser irradiation. Laser irradiation starts at *t*= 0 ps. From *t*= 10 ps to *t*= 40 ps, the region of high temperature expands because of the heat conduction from the irradiation spot to its surroundings. And at the same time, laser ablation is observed. The temperature of the target drops during 40 ps to 1460 ps. The cooling back to near initial temperatures, crystallization, all they are results of spreading of internal energy from hot spot by thermal conduction. The highest temperature drops below 102.0 K at 220 ps and 87.4 K at 440 ps in the simulation. By the end of the simulation *t*= 1460 ps, the temperature of the target is close to 60 K again. The temperature fluctuates corresponding to the stress wave. The temperature becomes higher when the stress wave begins to re-enter. A small region 20 nm×20 nm (x×z) in the radiation center (552.8, 752.8) (*x*, *z*) is chosen to explain this fluctuation. Fig. 5 shows the temperature and stress of this small region at the different times. An interesting phenomenon is observed in Fig. 5. The temperature goes up from 80 ps to 110 ps and from 150 ps to 180 ps where the re-entered stress wave causes the compressive stress increased, as the whole tendency of the temperature goes down. The atomic conduction cooling is the main source to let the temperature decrease. The small fluctuation means the acoustic heating also has effect on the temperature but much less than the conduction cooling. The fluctuation of temperature is due to the heating and cooling by the compression and rarefaction waves. The effect on the local temperature change due to the coupling effect between the temperature and the strain rate is proportional to the local strain rate. The results confirms the analytical solutions drawn by Wang [31] that the temperature change is related to stress ($\sigma$) as,

$$\frac{\partial T}{\partial t} \sim - \frac{B\beta_T T_0}{(B + \frac{4}{3}G)\ \rho C_p} \partial\sigma/\partial t \qquad (1)$$

Where, $G$ is shear modulus, $B$ is bulk modulus, $\beta_T$ is volumetric thermal expansion coefficient.

## 4. Phase Change Characteristics

Figure 6(a) shows the groups of atoms analyzed for the laser fluence of 10 J/m$^2$ at 1460 ps. According to Fig. 6(b), all the groups are back to solid phase at 1460 ps and will remain as solid. Their thermodynamic trajectories of pressure and temperature during the ablation process are shown in Fig. 6(b). The arrows indicate the progress of time. The phase boundary lines between solid, liquid and gas are taken from the standard argon diagrams of Encyclopedia Britannica Inc... From Fig. 6(b), it is seen that groups 1, 2, 3 and 4 experience material phase change. The pressure of the groups increase over the critical pressure suddenly. Groups 1, 2, 3 and 4 change from solid to liquid, then group 4 returns back to solid directly, and groups 1, 2 and 3 turn into supercritical fluid, finally return back to solid. On the other hand, group 5 which does not touch the phase boundary between solid and liquid, does not undergo phase change. When the laser irradiates the surface of the target, some groups of material are first raised to temperature higher than the critical temperature and become a super-critical fluid. After expansion, their temperature decreases, and they enter the unstable zone below the critical point as the phase separation occurs. When the laser irradiates the target inside, the thermodynamic trajectories of the groups suggest that the material can turn to supercritical fluid then return back to solid under this laser fluence, which is different from the laser ablation on the surface of target.

In the above discussion, it is important to note that local thermal equilibrium is achieved so that the temperature can be well defined. This can be verified by studying the velocity distribution at the locations of interest and comparing it with the Maxwell-Boltzmann distribution. In Fig. 7, the velocity distributions of atom group 2 in Fig. 6(a) are shown. The Maxwell-Boltzmann distributions that can best represent these velocity distributions are also shown. From Figure 7, it is seen that velocities of atoms indeed follow the equilibrium Maxwell-Boltzmann distribution before the laser radiation. During the 40 ps of laser heating process, the thermal equilibrium cannot be fully established, but after that velocities of atoms indeed follow the equilibrium Maxwell-Boltzmann distribution again. The Maxwellian distribution is fitted using both the temperature and velocity that subtracts the effect of the macroscopic velocity.

In this internal ablation, the whole process happens within an extremely confined domain. It will be interesting to study whether the phase state still follows that of normal Argon. Such study is conducted and summarized on group 2 in Table 1. In Table 1, $T_1$ is the temperature of normal argon defined by the pressure $P$ and specific volume $v$ which are given by the MD results and $T$ is calculated by the LAMMPS. The results indicate the status of the argon calculated by MD is almost consistent with parameters of the macroscopic status points.

An interesting thing is in the last line of Table1 at $t=$ 440 ps, from Fig. 8, the $P$-$T$ and $P$-$v$ diagrams at a number of time steps, it is solid near the separation line of solid and liquid. It does not follow the status of normal argon. When the $P$-$v$ diagram is used to show the status of argon, the boundary line between solid and solid-liquid is approximately by the line of $v=$ 24.72 m$^3$/mol, which can be calculated by the density of the solid argon. In initial state, the specific volume is 23.89 m$^3$/mol of the whole target, calculated by lattice constant= 0.5414 nm and 4 atoms in one lattice. But at the beginning of the laser radiation, the specific volume of the group 2 is about 26 m$^3$/mol, does not equal to 23.89 m$^3$/mol. The discrepancy is because the selected area is 2×2 nm, but the lattice constant is 0.5414 nm, so some atoms may not be included in the group, which makes the calculated value of specific volume higher than the actual specific volume. From Fig.8, the status point of 440 ps is near the boundary of solid and liquid, but in the side of solid on the $P$-$T$ diagram. Considering the discrepancy due to the chosen small area, the actual specific volume is lower than the calculated volume about 2~3 m$^3$/mol, the phase of the 440 ps point on the $P$-$v$ diagram is also in the side of solid, which matches the phase point on $P$-$T$ diagram. This also explains the discrepancy of the $T$ and $T_l$ in Table 1.

## 5. Crystallinity and orientation-radial distribution function（ODF）

In this work, crystallinity is used to judge the state of material at the atomic level. It is designed to show whether the material is close to the perfect crystal state or not, and the crystallinity is defined as [32]:

$$\phi\ r_{i,x}\ =\ \frac{1}{N}\ \sum_i e^{j2\pi(2r_{i,x}/\lambda)} \qquad (2)$$

Where, $r_{i,x}$ is the x coordinate of atom i. $\lambda$ is the light wavelength for crystallinity characterization assigned with the value of the lattice constant b = 5.414 Å. If atoms are regularly distributed in space with their spacing in the x direction equal to n(b/2), the function will be equal to 1. If the material is in amorphous state, the function will be much less than 1.

Figure 9 shows the contour of the crystallinity changes in time. Some representative time points are chosen to illustrate the change of the structure. The crystallinity's value reflects the structure of the material: value close to 1 indicating crystal structure close to the original state shown in t= 0 ps, and a low value indicating amorphous state. Compared with the stress distribution graph, it can be seen that when stress wave passes by, the crystallinity drops dramatically. This indicates the structure damage resulting from the high compressive stress. At 10 ps, the atoms in the irradiation zone experience a sudden temperature rise because of the high laser fluence. Stress wave travels in the materials and reaches the center again at t= 150 ps, and this makes the crystallinity worst at that moment. After the stress wave passes by and dissipates, the crystallinity of the damaged zone rises slowly and goes back to 1, which indicates that part of the material gets recrystallized after the ablation as shown in graph of 1460 ps.

Although the destruction of the crystal structure is illustrated by crystallinity function quite well, but there is still a problem whether the crystal is partially amorphous or just orientationally twisted. Under the influence of stress wave during the ablation process, material structure could be easily twisted and the crystallographic orientation could be changed. This makes the final structure of crystal significantly different from the initial crystallographic orientation variation in space. The orientation-radial distribution function (ODF) has been developed to illustrate this problem [32]. ODF provides information about the atom density and angle distribution of the crystal structure.

The 3D image of the face-centered cubic (FCC) structure of the argon crystal has been shown in Fig.10 (a), as it can be seen in the graph, atom 1, 2, 3 are the nearest atoms of the atom at origin. In this work, the atomic distance is calculated from the 3D space and angle is obtained from all atoms projected to the *X-Y* plane. At the initial state, the nearest atoms are 0, $\pi/4$, $\pi/2$, $3\pi/4$, $\pi$ in ideal situation.

It is noticed that there are some defects remaining after the cooling process in Fig.9. Three regions are chosen to illustrate the relation of ODF and defect in Fig.10 (b). Region 1 is the permanent defect and region 2 and 3 are the areas near region 1. The volumes of the three areas are the same, so they have almost the same amounts of atoms(about 58530±20) initially, and the atoms are all perfectly located in their positions according to FCC structure at the initial state as shown in Fig. 10 (c). ODF includes the information about the atom density variation in different angles. It can be seen that before the laser energy applied to the materials, the ODF of the crystal is exactly symmetrical. This indicates that the structure of the crystal which is perfect without any defect is with a large number of atoms distributing at certain radius (4.275Å, 6.525Å and 8.325Å). After the laser energy is applied to the material, the atoms are twisted to other angles as shown in ODF of 150 ps. In contrast with the temperature distribution, it can be seen that in the cooling process, atoms are relocated in its proper position, which would make partial defects to be repaired. From the Fig.10 (d) and (g), it can be clearly seen the angles of a large amount of atoms of region 1 are twisted and badly disorganized at 150 ps and 1460 ps. In contrast, the region 2 and region 3 just having some slightly defects finally. From Fig.10 (e)，(h)，(f)，and (i), atoms in these two regions almost perfectly locate around the specific angles at 1460 ps. This makes a good illustration that the phase of the target is to be solid finally but there still are twisted atoms to make permanent defect.

## 6. Conclusion

In this work, systematic atomistic modeling has been conducted to understand the phase change during laser internal ablation. The material destruction in the laser internal irradiation region is permanent. At the end of simulation at $t=$ 1460 ps, defects in the target are still observed. The propagation of stress wave and fluctuation of temperature are periodic in the recrystallization process. The initial peak temperature inside the target exceeds the critical temperature. This makes the solid argon the supercritical fluid, but the supercritical fluid argon returns to solid because of the limit of the stress and space, which is different from the ablation on the surface of material.

**Acknowledgement**

Support of this work by the National Science Foundation of China (No.51376164) and China Scholarship Council (No. 201406335016).

**List of figures and tables**

**Fig. 1** (a) Physical model for simulating the laser internal ablation in nanostructuring process. The laser energy is focused to a circular region with $R= 2$ nm in the target at $z= 81.15$ nm. The target measures 108.28 nm×10.82 nm×108.28 nm ($x \times y \times z$). (b) Laser beam intensity distribution. The full width at half maximum (FWHM) of the incident laser beam intensity distribution is 11.5 ps peaked at $t= 9$ ps. (c) Profile of the laser spot. The spot is round with a radius of $R= 2$ nm.

**Fig. 2** Snapshots of the case $E= 10$ J/m$^2$. Laser ablation starts from around $t= 5$ ps. Because of the internal ablation, the ablated material is prevented from moving freely in space. Recrystallization ($t= 200$-$1460$ ps) is observed following the ablation. Finally, there are defects in the final sample structure, as marked in the figure at $t= 1460$ ps.

**Fig. 3** Stress contour of the whole simulation system. A large compressive stress occurs inside the solid at the beginning of laser ablation. With the phase change and shock wave, the pressure drops down periodically. Approximately $t= 50$ ps, the stress wave reaches the $x$ boundary and is reflected back because of the confinement of the boundary. The reflected stress waves converge in the center of the $x$ direction at $t= 90$-$100$ ps, then the pressure drops down at 100 ps. This process of stress wave propagation is periodically with a period of about 80 ps.

**Fig. 4** Temperature contours of the case $E= 10$ J/m$^2$. The temperature of the area under direct laser irradiation goes up very quickly. And the high temperature results in ablation. Afterwards, the temperature in this area goes down in the recrystallization process. The highest temperature of the target reduce to 102.0K at $t= 220$ ps, 87.4 K at $t= 440$ ps and 68.0 K at $t= 1460$ ps.

**Fig. 5** Temperature and stress of the small region at different times. The small region measures 20 nm×20 nm ($x \times z$), centered at (552.8, 752.8) ($x, z$).

**Fig. 6** (a) Position of groups of atoms in the target. (b) Thermodynamic trajectories of groups of atoms at laser internal ablation in $P$-$T$ diagram. Black solid line is the phase boundary between liquid and gas; red solid line is the phase boundary between solid and liquid.

**Fig. 7** Velocity distributions for group 2 in Fig. 5. The solid line is the Maxwellian distribution. Red dots are MD simulation.

**Fig.8** Thermodynamic trajectories of groups of atoms at laser internal ablation in $P$-$T$ and $P$-$v$ diagrams. Black solid line is the phase boundary between liquid and gas; red solid line is the phase boundary between solid and liquid in the $P$-$T$ diagram. Black solid line is the phase boundary and red solid line is $T= T_c$ in the $P$-$v$ diagram.

**Fig.9** Crystallinity contours of the case $E= 10$ J/m$^2$. The whole domain is divided into cubes of size $1 \times 10.82 \times 1$ ($x \times y \times z$) nm$^3$. The melting area is characterized by low crystallinity (close to 0). The structure adjacent to the melting region is also destructed as shown in the whole process. In the recrystallization process, the laser irradiation spot has permanent defect characterized by the low crystallinity value in the figure at $t= 1460$ ps.

**Fig.10** Polar contours of the orientation-radial distribution function (ODF) of different regions. (a)The structure of FCC crystal in 3D space. The 1st order nearest atoms to atom 1 are atom 2, 3 and 4. The 1st nearest distance is 3.83 Å. (b) Location of the chosen region; (c) The ODF contour of the three region at initial state (Because the atoms are perfect crystal before the laser radiation, the ODF contours of the three regions are the same, so only one is chosen to illustrate); (d), (e), (f) ODF contours of the three regions at $t=$ 150 ps respectively; (g), (h), (i) ODF contours of the three regions at final state respectively.

**Table 1** Comparison of the temperature $T_1$ which is calculated by status software using parameters $P$ and $v$ and $T$ which are calculated by MD

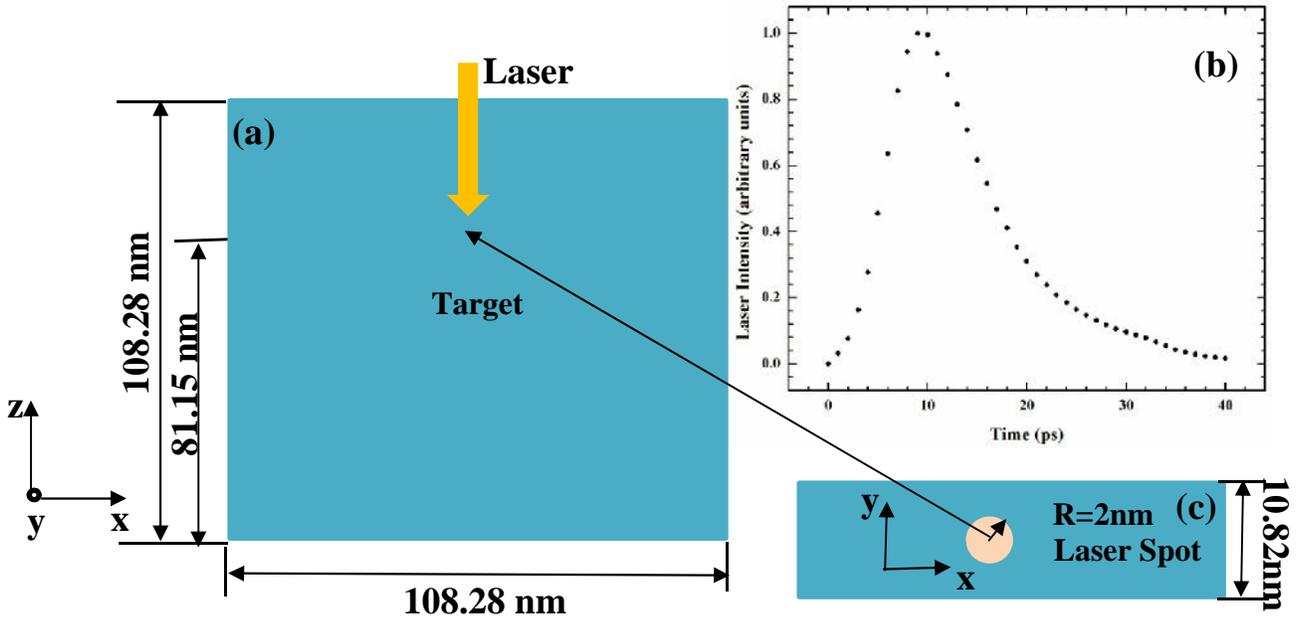

**Fig. 1** (a) Physical model for simulating the laser internal ablation in nanostructuring process. The laser energy is focused on a circular region with $R$= 2 nm in the target at $z$= 81.15 nm. The target measures 108.28 nm×10.82 nm×108.28 nm ($x×y×z$). (b) Laser beam intensity distribution. The full width at half maximum (FWHM) of the incident laser beam intensity distribution is 11.5 ps peaked at $t$= 9 ps. (c) Profile of the laser spot. The spot is round with a radius of $R$= 2 nm.

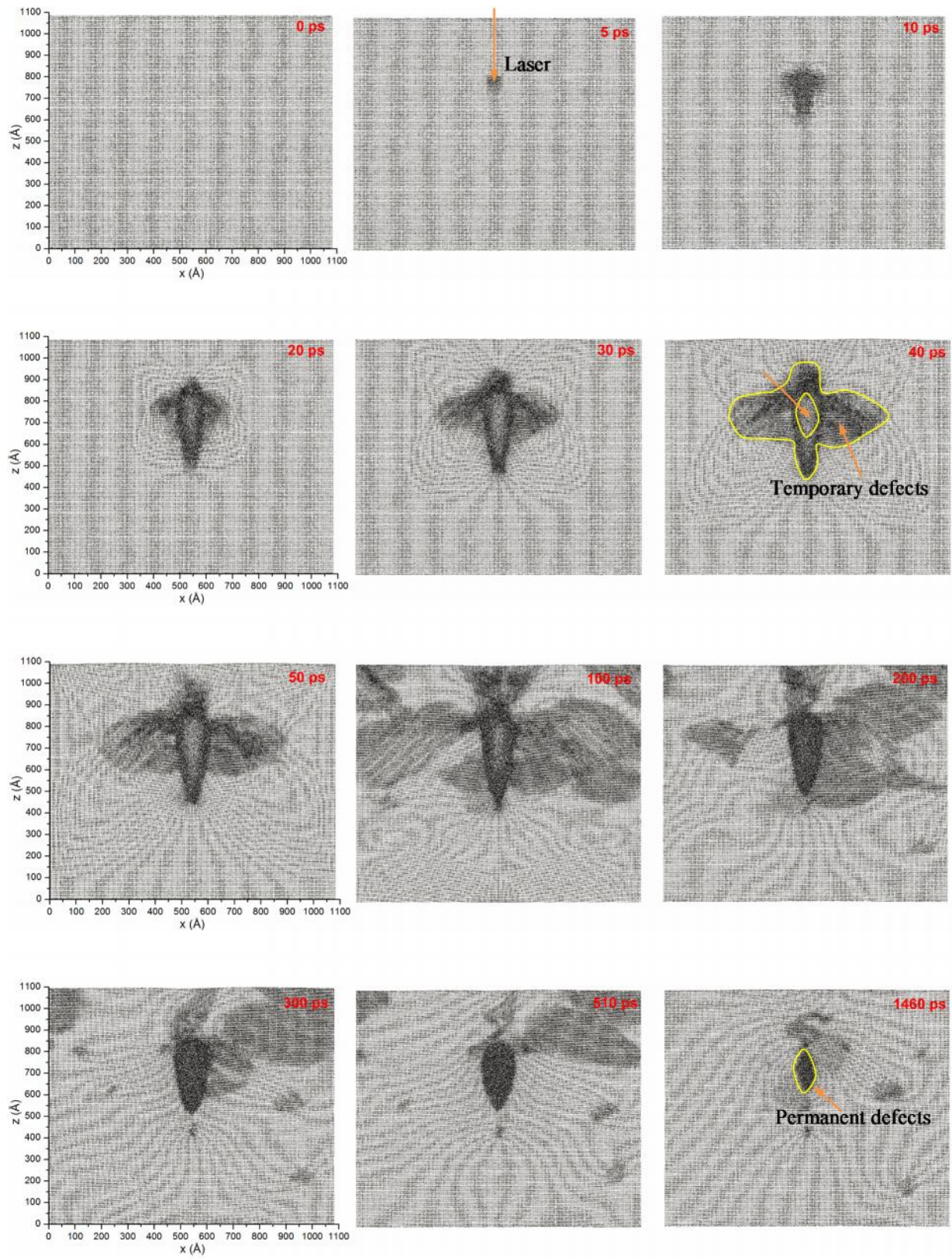

**Fig. 2** Snapshots of the case $E= 10$ J/m$^2$. Laser ablation starts from around $t= 5$ ps. Because of the internal ablation, the ablated material is prevented from moving freely in space. Recrystallization ($t= 200$-$1460$ ps) is observed following the ablation. Finally, there are defects in the final sample structure, as marked in the figure at $t= 1460$ ps.

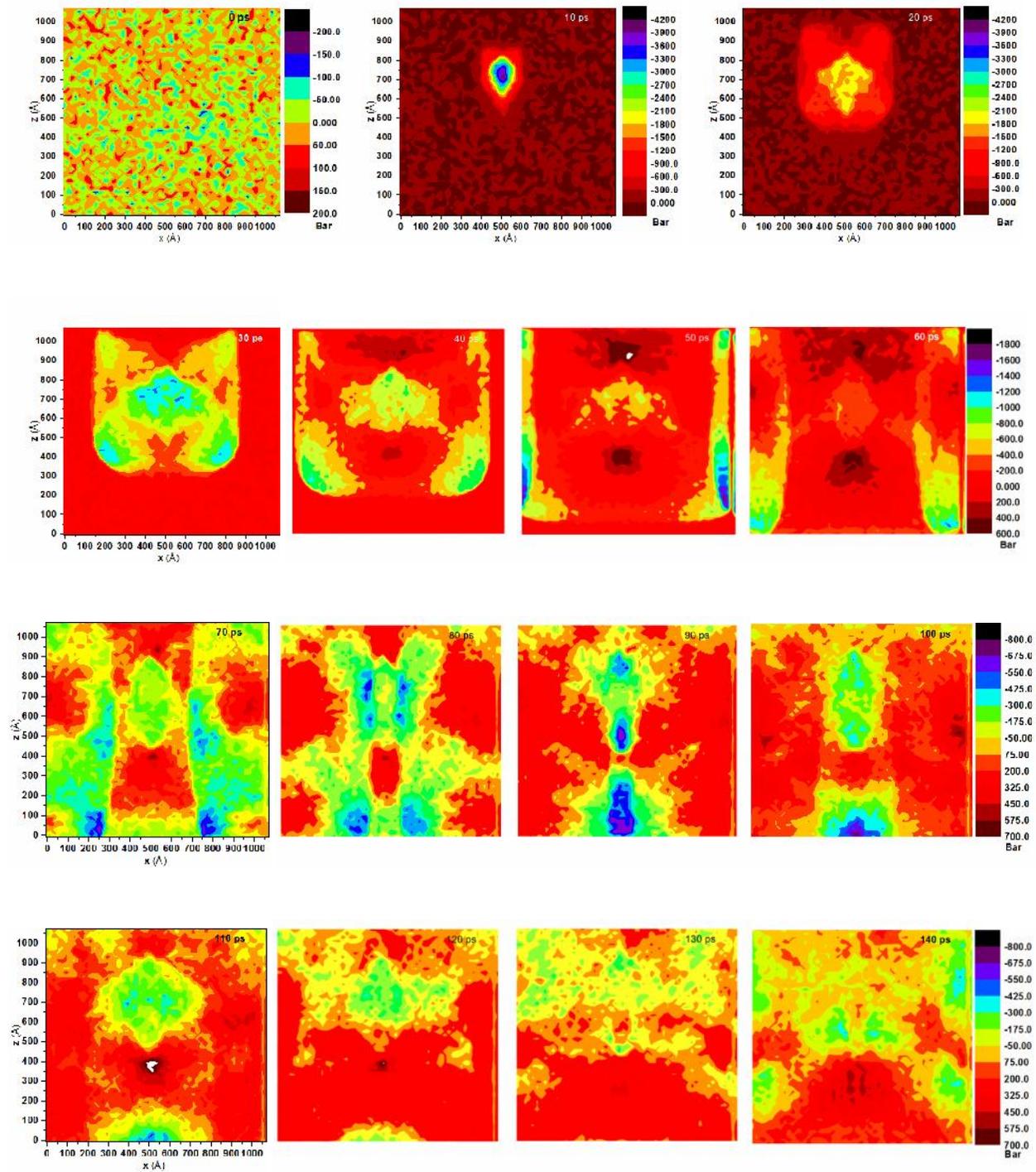

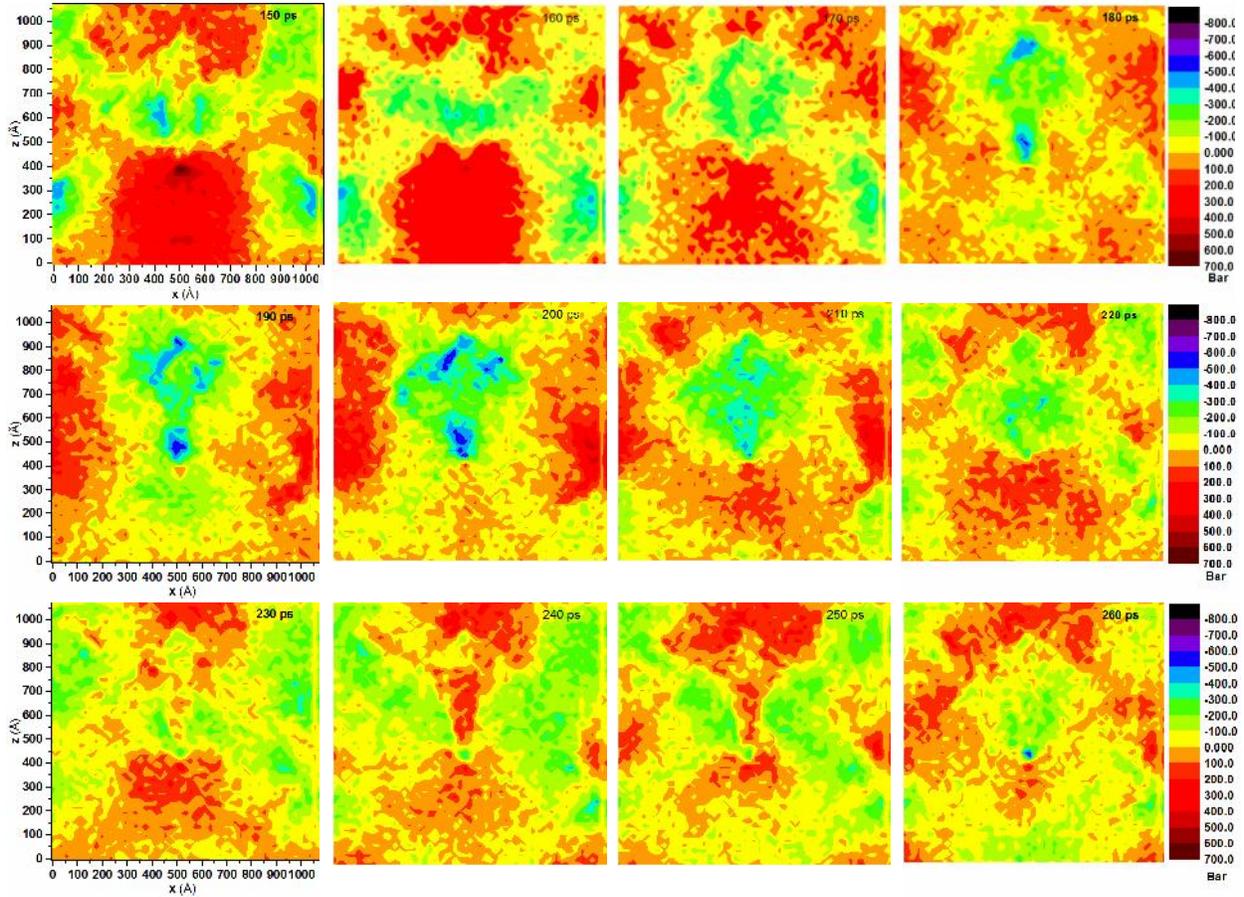

**Fig. 3** Stress contour of the whole simulation system. A large compressive stress occurs inside the solid at the beginning of laser ablation. With the phase change and shock wave, the pressure drops down periodically. Approximately $t=$ 50 ps, the stress wave reaches the $x$ boundary and is reflected back because of the confinement of the boundary. The reflected stress waves converge in the center of the $x$ direction at $t=$ 90-100 ps, then the pressure drops down at 100 ps. This process of stress wave propagation is periodically with a period of about 80 ps.

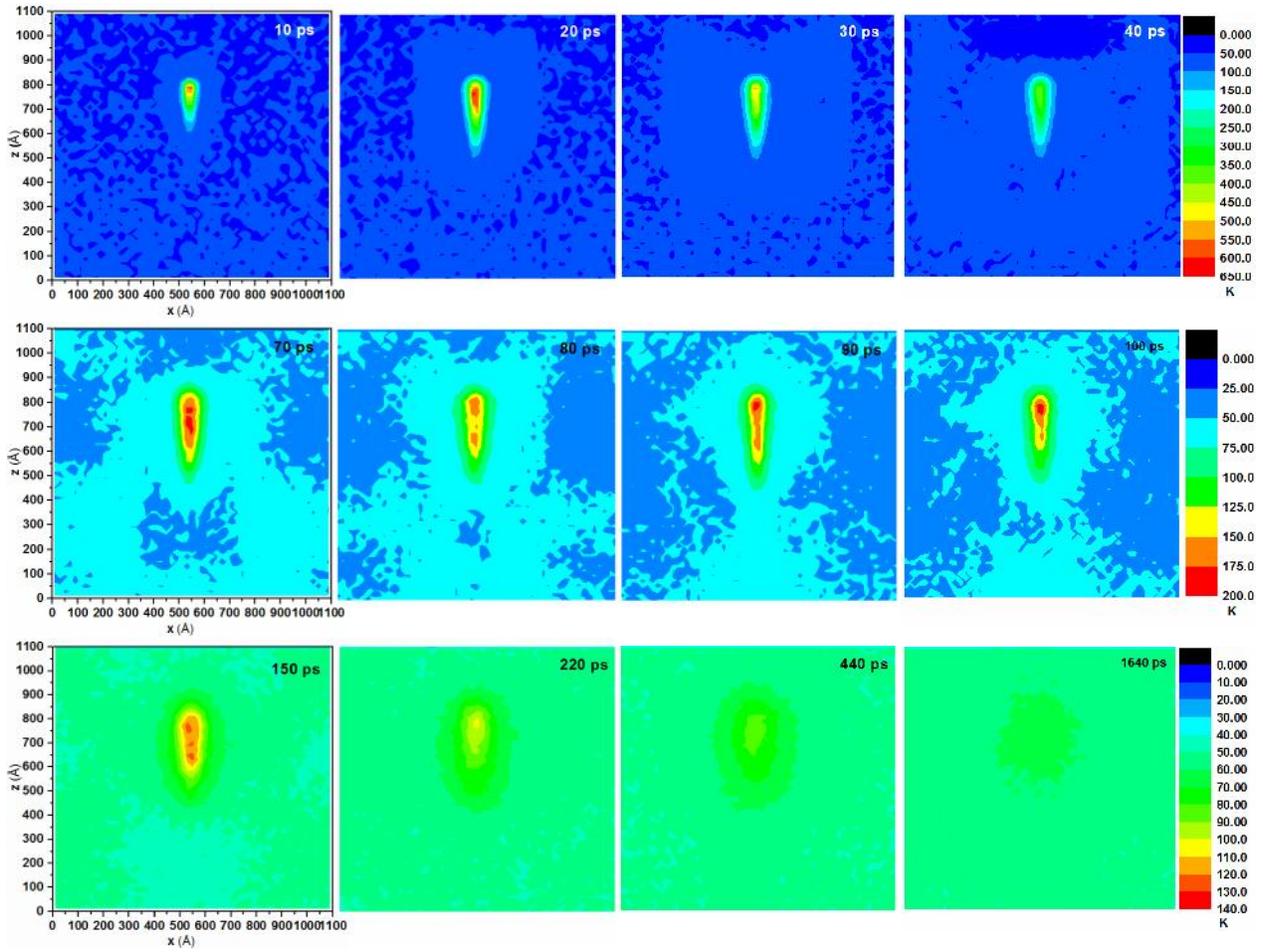

**Fig. 4** Temperature contours of the case $E$= 10 J/m$^2$. The temperature of the area under direct laser irradiation goes up very quickly. And the high temperature results in ablation. Afterwards, the temperature in this area goes down in the recrystallization process. The highest temperature of the target reduce to 102.0K at $t$= 220 ps, 87.4 K at $t$= 440 ps and 68.0 K at $t$= 1460 ps.

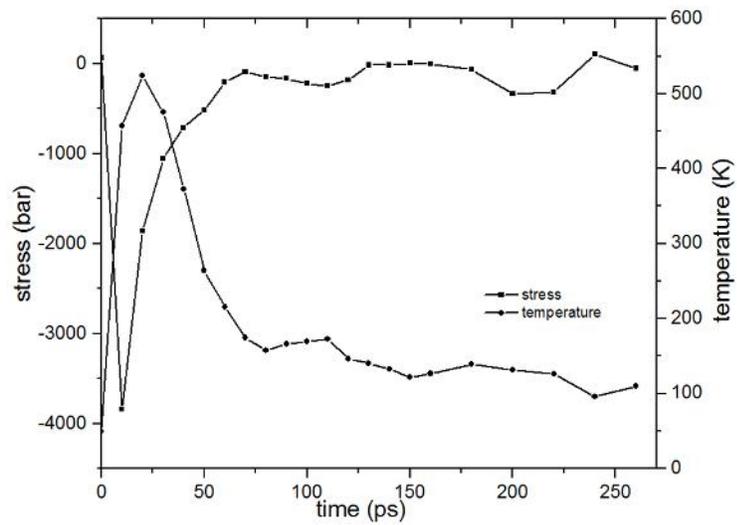

**Fig. 5** Temperature and stress of the small region at different times. The small region measures 20 nm×20 nm (*x*×*z*), centered at (552.8, 752.8) (*x*, *z*).

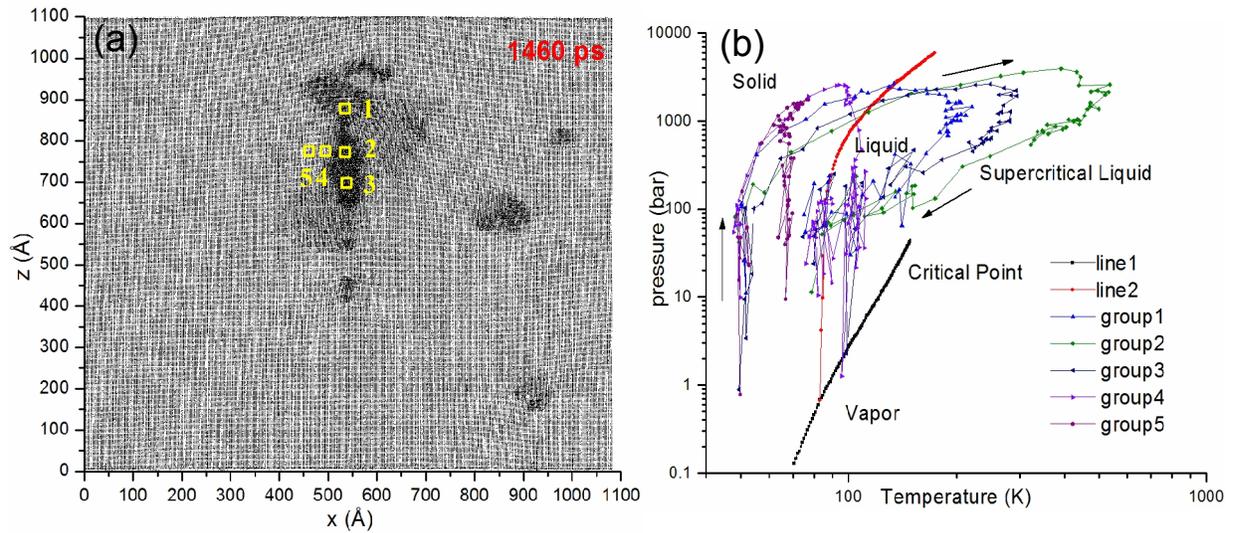

**Fig. 6** (a) Position of groups of atoms in the target. (b) Thermodynamic trajectories of groups of atoms at laser internal ablation in *P-T* diagram. Black solid line is the phase boundary between liquid and gas; red solid line is the phase boundary between solid and liquid.

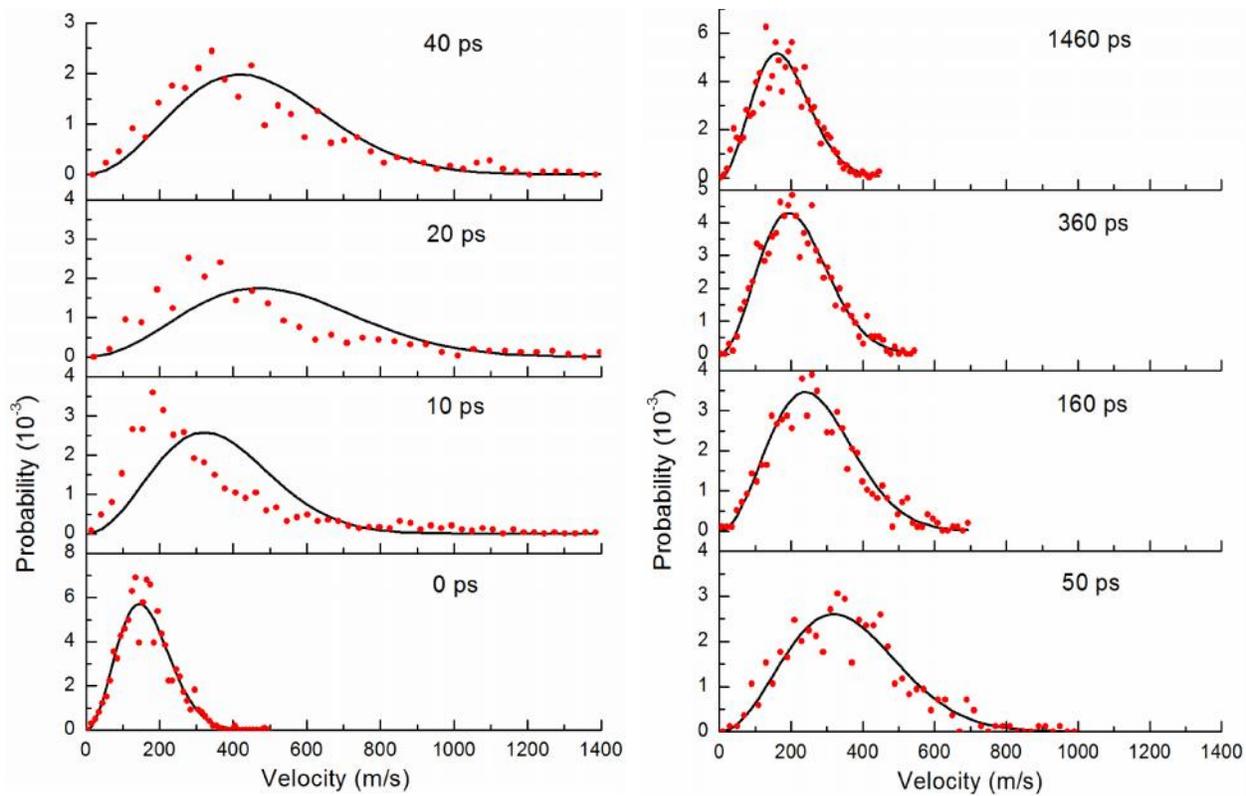

**Fig. 7** Velocity distributions for group 2 in Fig. 5. The solid line is the Maxwellian distribution. Red dots are MD simulation.

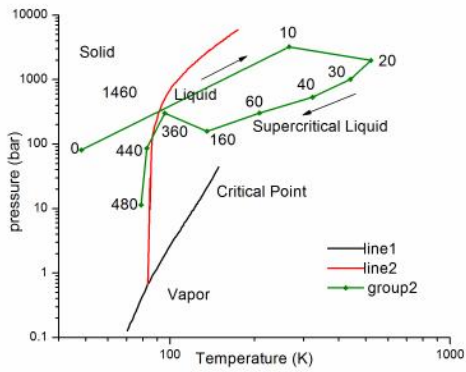 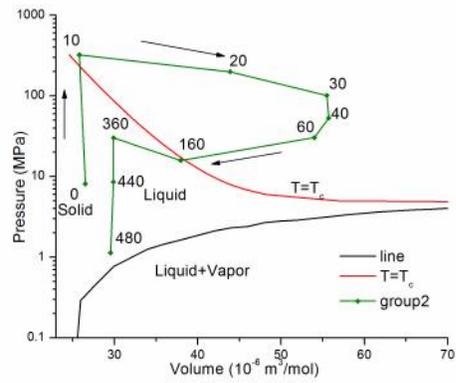

**Fig.8** Thermodynamic trajectories of groups of atoms at laser internal ablation in *P-T* and *P-v* diagrams. Black solid line is the phase boundary between liquid and gas; red solid line is the phase boundary between solid and liquid in *P-T* diagram. Black solid line is the phase boundary and red solid line is $T = T_c$ in the *P-v* diagram.

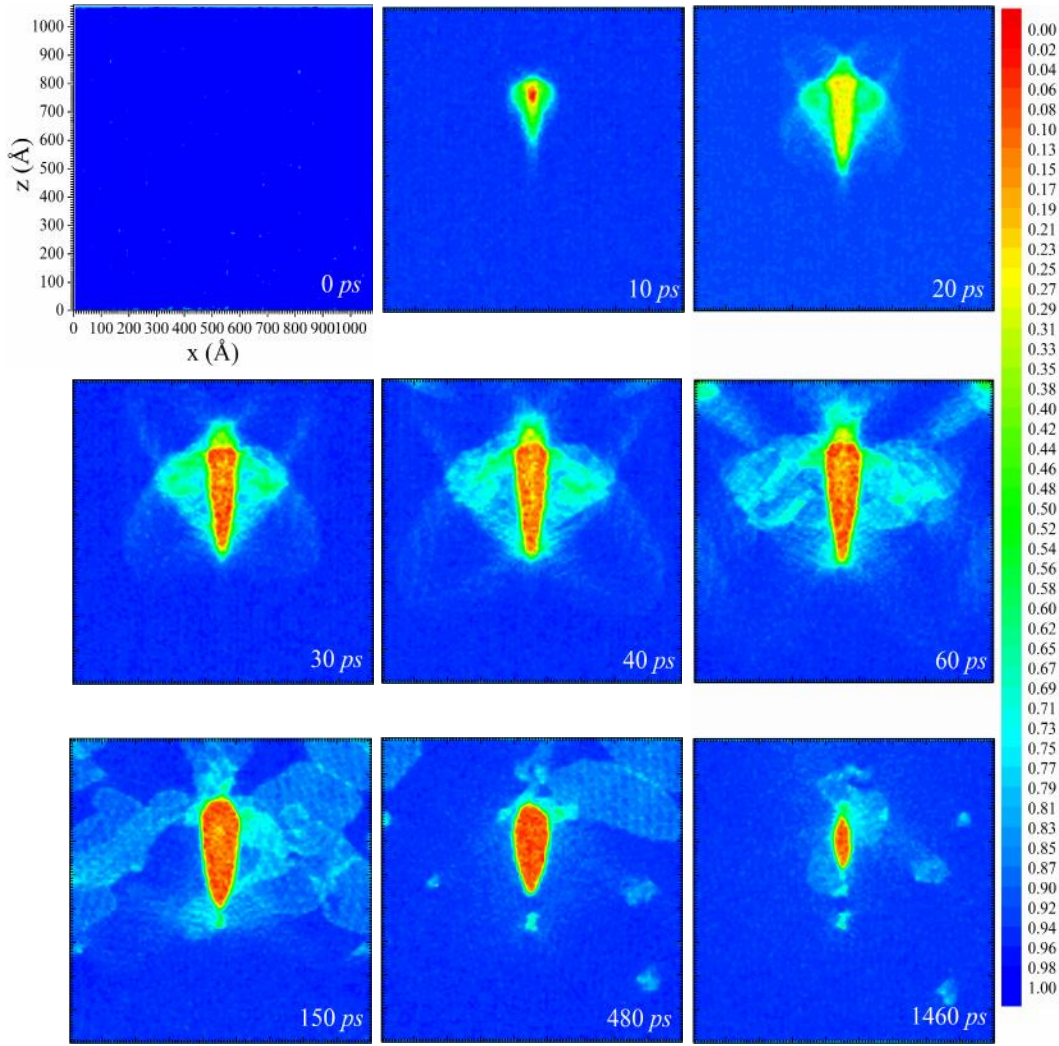

**Fig.9** Crystallinity contours of the case $E= 10$ J/m$^2$. The whole domain is divided into cubes of size $1\times10.82 \times1$ ($x\times y\times z$) nm$^3$. The amorphous area is characterized by low crystallinity (close to 0). The structure adjacent to the amorphous region is also destructed as shown in the whole process. In the recrystallization process, the laser irradiation spot has permanent defect characterized by the low crystallinity value in the figure at $t= 1460$ ps.

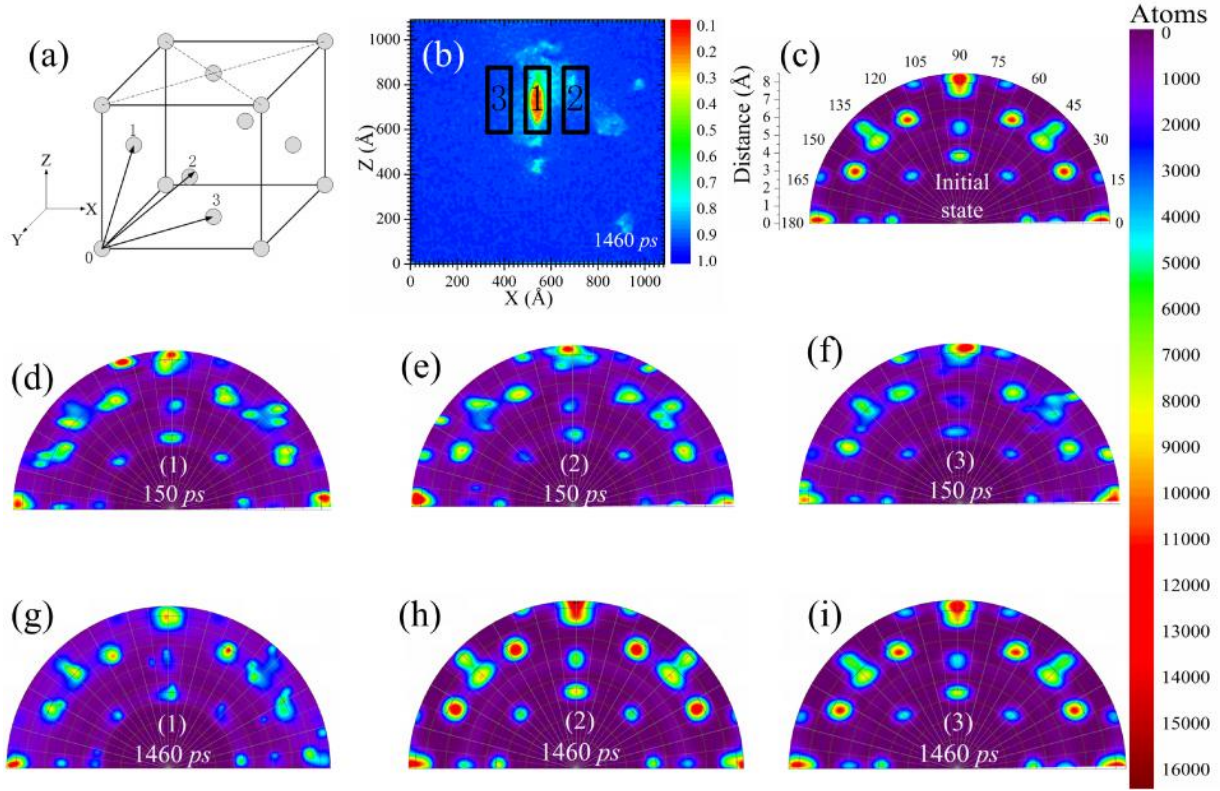

**Fig.10** Polar contours of the orientation-radial distribution function (ODF) of different regions. (a) The structure of FCC crystal in 3D space. The 1st order nearest atoms to the atom 0 are atom 1, 2 and 3. The 1st nearest distance is 3.83 Å. (b) Location of the chosen region; (c) The ODF contour of the three region at initial state (Because the atoms are perfect crystal before the laser radiation, the ODF contours of the three regions are the same, so only one is chosen to illustrate); (d), (e), (f) ODF contours of the three regions at $t$ =150 ps respectively; (g), (h), (i) ODF contours of the three regions at final state respectively.

**Table 1** Compare of the temperature $T_1$ which is calculated by status software using parameters $P$ and $v$ and $T$ which is calculated by MD

| Time (ps) | P (Mpa) | V ($10^6$ m³/mol) | $T_1$ (K) | T (K) |
|---|---|---|---|---|
| 10 | 322.7 | 25.79 | 197 | 266 |
| 20 | 199.3 | 43.89 | 539 | 521 |
| 30 | 101.3 | 55.48 | 455 | 440 |
| 40 | 53.5 | 55.71 | 303 | 322 |
| 60 | 30.4 | 53.98 | 224 | 208 |
| 160 | 15.9 | 37.95 | 149 | 135 |
| 360 | 30.2 | 29.90 | 115 | 96 |
| 440 | 8.6 | 29.87 | 102 | 82 |